\documentclass[prb,twocolumn,showpacs,amsmath]{revtex4}

\usepackage{graphicx}
\begin{document} 
\title{Two Distinct Electronic Contributions in the Fully Symmetric Raman 
Response of High $T_{c}$ Cuprates}.

\author{M. Le Tacon$^{1,2}$, A. Sacuto$^{1,2}$, and D. Colson$^{3}$}
\address{$^{1}$Laboratoire de Physique du Solide ESPCI, 10 rue Vauquelin, 75231 Paris, France\\
$^{2}$Mat\'eriaux et ph\'enom$\grave{e}$nes Quantiques (UMR 2437 CNRS), Universit\'e Paris 7, 2 place Jussieu %
75251 Paris, France\\ 
$^{3}$Service de Physique de l'Etat Condens\'{e}e, CEA-Saclay, 91191 Gif-sur-Yvette, France}

\begin{abstract} 
We show by resonance effects in HgBa$_2$CuO$_{4+\delta}$ (Hg-1201) and by Zn substitutions in 
YBa$_2$Cu$_3$O$_{7-\delta}$ (Y-123) compounds that the fully symmetric Raman spectrum has 
two distinct electronic contributions. The A$_{1g}$ response consists of  
the superconducting pair breaking peak at the 2$\Delta $ energy and a 
collective mode close to the magnetic resonance energy. These experimental 
results reconcile the \textit{d-wave} model to the A$_{1g}$ Raman response function in so 
far as a collective mode that is distinct from the pair breaking peak is present in the 
A$_{1g}$ channel.
\end{abstract}

\maketitle
\date{\today} 

\noindent \qquad In the last few years, it has been well established that the superconducting gap 
of the hole-doped cuprates at the optimal doping regime has the 
$d_{x^2-y^2}$ symmetry \cite{Dev94,Kang96,Sacuto00}. 

This symmetry manifests itself in the low energy part of the Raman spectra.
In the B$_{2g}$ channel \cite{polar} (probing the nodal directions), the electronic 
continuum behaves as a linear function of the Raman shift, while it follows 
a cubic law in the B$_{1g}$ channel \cite{polar} (anti-nodal directions)(see ref. \onlinecite{Dev95}).  
In the latter one, a well defined pair breaking peak near 2$\Delta$ = 8$k_{B}T_c$
is observed. 
However, existing theories based on the $d_{x^2-y^2}$ model, fail to 
reproduce the position, the intensity, and the shape of the broad electronic 
peak observed in the fully symmetric A$_{1g}$ channel \cite{polar,Sacuto00,Wenger97,Strohm97}. 
Expansion of the Raman vertex to the second order of the Fermi surface 
harmonics \cite{Dev95} and resonant effects \cite{Sherman02} have been proposed 
to reproduce the relative A$_{1g}$ peak position and intensity with respect to 
that of B$_{1g}$. In these pictures, the A$_{1g}$ peak is treated as another 
manifestation of the pair breaking peak observed in the B$_{1g}$ channel. 
Unfortunately, the back flow prevents the reproduction 
of the location, on one hand, and on the other hand, of the sharpness and the strong intensity 
of the A$_{1g}$ peak. For a 
generic tight-binding model, the calculated screened A$_{1g}$ channel is only a 
tiny fraction of the B$_{1g}$ response \cite{Wenger97}. This is in clear 
contradiction to all experiments and most studies showing the magnitude of 
the A$_{1g}$ peak being even larger than the B$_{1g}$ \cite{Gallais02,Chen94, 
Gasparov97, Sacuto98}.
 In this paper we show that the A$_{1g}$ response has two components: one component 
originating from the pair breaking close to the 2$\Delta$ energy 
and the other from a collective mode which tracks the magnetic resonance 
\cite{Gallais02,MLT}. In this sense, our experimental results reconcile the A$_{1g}$ Raman 
response of the cuprates at the optimal doping regime with the \textit{d-wave} model  in
so far as a collective mode is present in the A$_{1g}$ channel.

Electronic Raman Scattering (ERS) measurements have been carried out with a 
JY T64000 triple spectrometer in subtractive configuration using different 
lines of mixed Argon - Krypton laser gas. The Raman spectra were corrected 
for the spectrometer response, the Bose factor and the optical constants 
producing the imaginary part $\chi^{\prime\prime}(\omega)$ of the Raman response. 
The crystals were mounted in vacuum (10$^{-6}$ mbar) on the cold finger of a liquid-helium flow
cryostat. The power density was about 10W/cm$^{2}$ on the sample surface, and the laser spot heating 
estimated from the Anti-Stokes/Stokes intensity ratio of the Raman responses was less 
than 3K. 

\begin{figure}[tbh]
\begin{center}\hspace{-10mm}
\includegraphics[width=9.5cm]{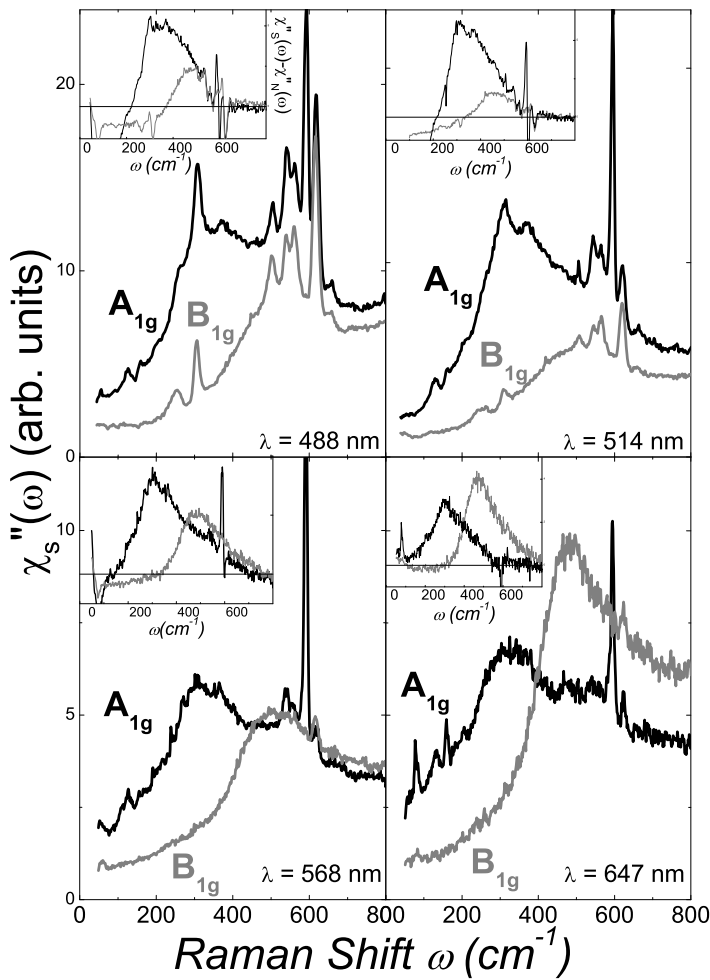}
\end{center}\vspace{-5mm}
\caption{Raman responses  $\chi^{\prime\prime}_{S}(\omega)$
 of optimally doped Hg-1201 for different excitations lines in the A$_{1g}$ (black line)
 and B$_{1g}$ (gray line) channels. 
The insets exhibit $\chi^{\prime\prime}_{S}(\omega)$-
$\chi^{\prime\prime}_{N}(\omega)$ for both A$_{1g}$ and B$_{1g}$ channels.}
\label{fig:plane}
\end{figure}

Let's focus first on ERS measurements of optimally doped Hg1201 single crystals ($T_c$=95K).
They have been grown by flux method whose detailed procedure is described elsewhere 
\cite{Colson94}.
Figure \ref{fig:plane} shows the superconducting Raman responses $\chi^{\prime\prime}_{S}(\omega)$ 
of Hg-1201 obtained for various excitations lines in A$_{1g}$ and B$_{1g}$ channels.

 The Raman responses are composed of a broad electronic continuum surrounded by 
an assembly of narrow peaks corresponding to the well identified 
phonons \cite{Krantz94}.
At first glance, the Raman responses for each excitation line (E.L.) reveal that 
the A$_{1g}$ continuum exhibits a strong maximum around 330 cm$^{-1}$, with an 
asymmetric part in its high energy side. This manifests itself as
a bump for blue (488 nm) and green (514 nm) E.L., and as a ``plateau'' 
for yellow (568 nm) and red (647 nm) ones, which  are around 520 cm$^{-1}$
near the maximum of the B$_{1g}$ continuum that corresponds to the pair breaking peak.  

The Raman responses of the blue and green lines show strong phonon features super-imposed 
to the electronic continuum near 520 cm$^{-1}$ which complicates the extraction of the 
electronic background.
 On the contrary, under the yellow and red E.L., the phonon modes are out of resonance 
thus their structures are strongly reduced and the electronic contribution can be easily extracted. 
Subtractions of the normal $\chi^{\prime\prime}_{N}(\omega)$ response from the superconducting 
$\chi^{\prime\prime}_{S}(\omega)$ one are reported in the insets of Figure 1. 
The Raman responses $\chi^{\prime\prime}_{S}(\omega)$-$\chi^{\prime\prime}_{N}(\omega)$
for the yellow and red lines are almost free of phonon contribution. 
The broad continua in the A$_{1g}$ and B$_{1g}$ channels correspond to the electronic contributions 
from the superconducting state, and the sharp features show misfits between the superconducting and 
normal phonon structures.
After substraction of the normal state contribution, the A$_{1g}$ response is still asymmetric, and for 
each E.L., the high energy part of this response is centered near the maximum of the B$_{1g}$ superconducting
gap. The asymmetry of the A$_{1g}$ response is thus intrinsic to the superconducting state.

To go further and prove that the broad A$_{1g}$ peak consists effectively of two distinct 
electronic components, we have performed ERS measurements on high quality 
optimally doped YBCO single crystals grown by the self flux method 
\cite{Kaiser87}, where copper is substituted by zinc. Zn is a divalent ion 
known to substitute preferentially in the CuO$_{2}$ layers without altering 
the carrier concentration \cite{Bobroff99}. In addition to the pure 
YBa$_{2}$Cu$_{3}$O$_{7-\delta }$(Y-123, \textit{T}$_{c}$= 92K), we have studied 
YBa$_{2}$(Cu$_{1-y}$Zn$_{y})_{3}$O$_{7-\delta }$ single crystals with y=0.005 
(\textit{T}$_{c}$=87K), y=0.01 (\textit{T}$_{c}$=83K), y=0.02 (\textit{T}$_{c}$=73K) and y=0.03 (\textit{T}$_{c}$= 64K).
 Zn concentration was verified by chemical analysis using an electron probe. $T_c$
measurements were obtained from DC- magnetization and we found $dT_c/dy \sim$ - 10K/{\%}.
 
\begin{figure}[tbh]
\begin{center}\hspace{-9.3mm}
\includegraphics[width=9.5cm,height=15.5cm]{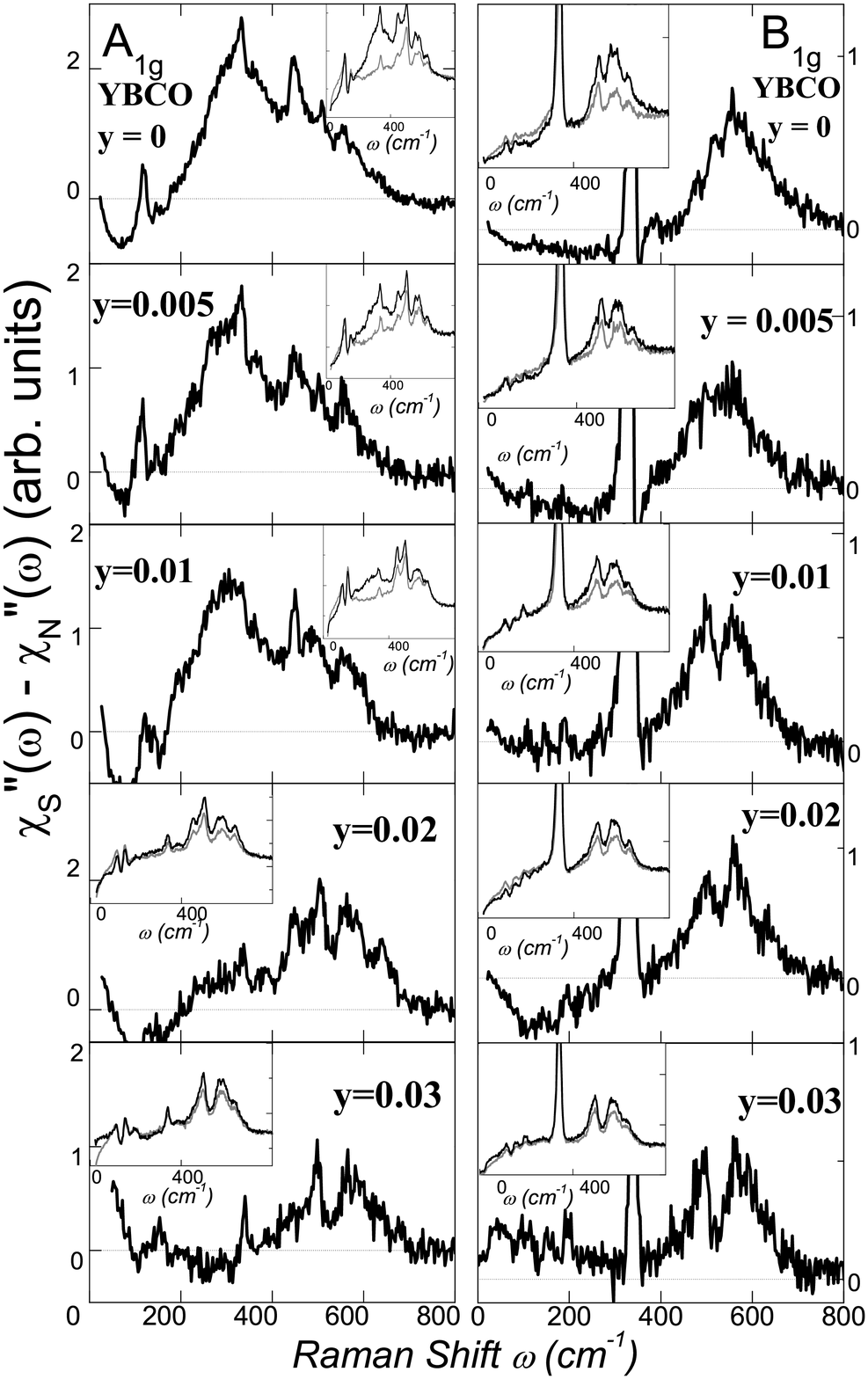}
\end{center}\vspace{-5mm}
\caption{$\chi^{\prime\prime}_{S}(\omega)-\chi^{\prime\prime}_{N}(\omega)$
A$_{1g}$ (left panel) and B$_{1g}$ (right panel) Raman responses of optimally doped 
YBa$_{2}$(Cu$_{1-y}$Zn$_{y})_{3}$O$_{7-\delta }$ for various Zn concentrations y. $\lambda$=514nm.
In each inset are plotted the Raman response functions in the normal (gray line) and superconducting
(black line) states.}
\label{fig:Zn}
\end{figure}

Figure \ref{fig:Zn} shows the $\chi^{\prime\prime}_{S}(\omega)$-$\chi^{\prime\prime}_{N}(\omega)$
 Raman responses in A$_{1g}$ and B$_{1g}$ channels in Y-123 for various Zn contents. 
Insets exhibit the A$_{1g}$ and B$_{1g}$ Raman 
responses in the normal and superconducting states before subtraction. 
The A$_{1g}$ and B$_{1g}$ Raman responses show a set of sharp phonon peaks lying on a strong
electronic background. In the A$_{1g}$ channel, for pure 
YBCO, the $\chi^{\prime\prime}_{S}(\omega)$-$\chi^{\prime\prime}_{N}(\omega)$ Raman 
response shows a broad and strong asymmetric peak 
which spreads out in the high energy side and reaches its maximum close to 
330 cm$^{-1}$. In the B$_{1g}$ channel, the $\chi^{\prime\prime}_{S}(\omega)$-$\chi^{\prime\prime}_{N}
(\omega)$ response for pure YBCO, exhibits a well defined and nearly symmetric peak close to 530 cm$^{-1}$.
 These A$_{1g}$ and B$_{1g}$ superconducting spectra are very similar to those 
obtained from Hg-1201 at the optimal doping. Here again, the A$_{1g}$ response exhibits a 
maximum around 330 cm$^{-1}$ with an asymmetric part which extends up to the pair breaking peak 
near 530 cm$^{-1}$. The positions of the A$_{1g}$ and B$_{1g}$ peaks are nearly the same 
for both Y-123 and Hg-1201. The changes in the band structure induced by two CuO$_{2}$ 
layers instead of one CuO$_{2}$ layer do not affect the A$_{1g}$ and B$_{1g}$ peak 
positions rather, the critical temperature (92 K for Y-123 and 95 K for 
Hg-1201) seems to govern the A$_{1g}$ and B$_{1g}$ peak energies at the optimal 
doping. This is observed for many cuprates where the B$_{1g}$ peak is found 
close to 8$k_{B}T_c$ and the A$_{1g}$ peak maximum close to 5$k_{B}T_c$
at the optimal doping (see Table 1 of Ref. \onlinecite{Gallais02,MLT}). 
Adding some Zn in the pure Y-123, one can see that the intensity of the pair breaking peak seen in the B$_{1g}$ 
channel decreases, but the peak does not disappear and is still present even in the sample with y=0.03 ($T_c$ =63K),
contrary to what is suggested in Ref. \onlinecite{Martinho}.
The B$_{1g}$ peak does not shift in energy, and thus does not follow $T_c$, but this effect and other
related to nonmagnetic impurity substitutions in YBCO will be discuss in a next paper. 

Let's focus now on the A$_{1g}$ channel. As Zn content increases, the low energy 
contribution in the A$_{1g}$ response becomes broader, shifts to a lower energy 
(from 330 to 300 cm$^{-1}$), strongly decreases in its intensity, and finally 
disappears. On the contrary, the high energy contribution in the A$_{1g}$ 
response moderately decreases in its intensity but keeps the same position 
at 530 cm$^{-1}$. 

For y=0.01, the $\chi^{\prime\prime}_{S}(\omega)$-$\chi^{\prime\prime}_{N}(\omega)$ 
A$_{1g}$ response, clearly shows two components. The first is centred 
at 300 cm$^{-1}$ and the second is close to 530 cm$^{-1}$.
For higher Zn concentrations, the intensity ratio of the upper and the 
lower energy parts of the A$_{1g}$ response, is reversed in such a way that for 
y=0.03, the lower energy component completely disappears while the upper 
component persists. A straightforward comparison between the left and right 
panels reveals that the high energy component (530 cm$^{-1}$) in the A$_{1g}$ response 
tracks the B$_{1g}$ peak. 
For both y=0.02 and y=0.03 when the lower energy component in the A$_{1g}$ 
response is weak and does no longer mix with the higher energy component, the 
ratio of the spectral weight between the higher energy component in the A$_{1g}$ 
response and the pair breaking peak in the B$_{1g}$ response remains constant. 
In these cases the peaks observed in A$_{1g}$ and B$_{1g}$ channels are located at the same energy 
\cite{com2} and correspond both to the pair breaking peak. 

This gives experimental evidence that the A$_{1g}$ response has two 
distinct components and that the one of higher energy corresponds to the pair breaking peak. 
As the B$_{1g}$ one probes the anti-nodal directions of the $d_{x^2-y^2}$ superconducting gap, 
and the A$_{1g}$ Raman response has no symmetry restriction, it is therefore not 
surprising to observe the pair breaking peak in both A$_{1g}$ and B$_{1g}$ channels. 
The low energy component of the A$_{1g}$ Raman response corresponding to the maximum of the electronic
 continuum is intrinsic to the superconducting state and disappears above $T_c$ as it was already pointed
 out in previous works \cite{Gallais02,MLT}. The A$_{1g}$ peak is located at 5$k_BT_c$ well below the 2$\Delta$ 
energy gap (8$k_BT_c$) and therefore cannot be induced by individual electronic excitations which required 
energies beyond 2$\Delta$. 
As a consequence the A$_{1g}$ mode has to be a bound state of quasi-particle pairs at an energy less than 2$\Delta$ 
and refers to a collective mode. We have not yet identified the origin of the A$_{1g}$ mode but several scenarios
 can be figured out. Among them the Bogoliubov-Anderson collective mode \cite{Anderson} calculated in the frame work of 
the \textit{d-wave}  model merits to be considered as well as a double magnon with a zero spin flip energy as 
suggested by E. Demler \cite{Demler}. Zero spin flip energy is possible for a \textit{d-wave} superconductor if we 
consider spin flip excitations over two nodal regions.  


Zero spin flip quasi-particle excitations have already been invoked to explain the quadratic increase of the $^{17}$O 
spin-lattice relaxation rate under magnetic fields accross the vortex lattice NMR spectrum in YBCO \cite{Vesna}. 
In our case, the A$_{1g}$ peak tracks the acoustic magnetic resonance detected by inelastic neutron scattering 
\cite{Rossat91} at \textbf{Q}=($\pi,\pi$) as previously shown \cite{Gallais02,MLT}.
 A double spin flip of transfered momenta \textbf{Q}=($\pi,\pi$) and \textbf{Q}=($-\pi,-\pi$) is then needed for 
preserving the total transfer momentum close to zero in the Raman scattering process.
The first spin flip is over two anti-nodal  regions and costs the magnetic resonance energy  whereas  
the second spin flip over two nodal regions costs zero energy. 
In this Raman process the A$_{1g}$ mode takes the same energy as the magnetic resonance as expected experimentally. 
Theoretical investigations on this last scenario are in progress.

In summary, the ERS spectra of Hg-1201, free of phonon peaks, reveal that the 
A$_{1g}$ response and its asymmetry near the pair breaking peak are of electronic 
origin. Moreover, ERS in Y-123 substituted with Zn shows that the A$_{1g}$ peak has two distinct 
components: one at the higher energy corresponding to the pair breaking peak 
observed in the B$_{1g}$ channel and the other at lower energy corresponding to another 
electronic contribution that is distinct from the pair breaking peak. This 
study reconciles the A$_{1g}$ Raman response function with the \textit{d-wave} model where the pair 
breaking peak manifests itself in both B$_{1g}$ and A$_{1g}$ channels. This implies 
the existence of a charge collective mode below the pair breaking peak 
energy which we have previously related to the magnetic resonance. 

\bigskip
\textbf{Acknowledgements}
We wish to thank M. Cazayous, Y. Gallais, V.F. Mitrovic, V.N. Muthukumar, E.Demler, A. Benlagra and S. Nakamae 
for very fruitful discussions.


\bigskip

\begin{references} 

\bibitem{Dev94} T. P. Devereaux, D. Einzel, B. Stadlober, R. Hackl, D.H. Leach, and J.J. 
Neumeier, Phys. Rev. Lett. \textbf{72}, 396 (1994).

\bibitem{Kang96} M. Kang, G. Blumberg, M. V. Klein, and N. N. Kolesnikov, Phys. Rev. Lett. 
\textbf{77}, 4434 (1996).

\bibitem{Sacuto00} A. Sacuto, J. Cayssol, Ph. Monod, and D. Colson Phys. Rev.B \textbf{61}
, 7122 (2000).

\bibitem{polar} The B$_{2g}$ and B$_{1g}$ channels are obtained for cross polarizations for incident and
scattered light, where the incident electric field direction is along and at 45 degrees of the Cu-O bounds
of CuO2 layers (i.e. the \textbf{a} and \textbf{b} crystal axes) respectively.
Parallel polarizations at 45 degrees of the \textbf{a} and \textbf{b} crystal axes give access to the A$_{1g}$+B$_{2g}$
 spectrum, pure A$_{1g}$ channel is obtained by subtracting the B$_{2g}$ spectrum to the A$_{1g}$+B$_{2g}$ one. 

\bibitem{Dev95} T. P. Devereaux and D. Einzel, Phys. Rev. B \textbf{51}, 16336 (1995);
Phys. Rev. B \textbf{54}, 15547 (1996)

\bibitem{Wenger97} F. Wenger and M.K\"{a}ll, Phys. Rev. B \textbf{55}, 97 (1997)

\bibitem{Strohm97} T. Strohm and M.Cardona, Phys. Rev. B \textbf{55}, 12725 (1997)

\bibitem{Sherman02} E. Ya. Sherman, C. Ambrosch-Draxl, and O. V. Misochko, Phys. Rev. B \textbf{65}, 
140510 (R) 2002. 

\bibitem{Gallais02} Y. Gallais, A. Sacuto, Ph. Bourges, Y. Sidis, A. Forget, and D. Colson, 
Phys. Rev. Lett. \textbf{88}, 177401 (2002), and ref. therein.


\bibitem{MLT} M. Le Tacon, Y. Gallais, A. Sacuto, and D. Colson, Journal of Physics and Chemistry of Solids, 
SNS2004 proceedings, \textit{to be published}.

\bibitem{Chen94} X. K. Chen, J.C. Irwin, H.J. Trodhal, T. Kimura, and K. Kishio, Phys. Rev. 
Lett. \textbf{73}, 3290 (1994).

\bibitem{Gasparov97} L.V. Gasparov, P. Lemmens, M. Brinkmann, N. N. Kolesnikov, and 
G.G\"{u}ntherodt, Phys. Rev B \textbf{55}, 1223 (1997)

\bibitem{Sacuto98} A. Sacuto, R. Combescot, N. Bontemps, C. A. M\"{u}ller, V. Viallet, and 
D. Colson, Phys. Rev. B \textbf{58}, 11721 (1998). 

\bibitem{Colson94} D. Colson, A. Bertinotti, J. Hammann, J. F. Marucco, and A. Pinatel, Physica 
C \textbf{233}, 231, (1994). 

\bibitem{Krantz94} M.C. Krantz, C. Thomsen, H.J. Mattausch, and M. Cardona, Phys. Rev. B 
\textbf{50}, 1165, (1994).

\bibitem{Kaiser87} D. L. Kaiser, F. Holtzberg, B. A. Scott, and T.R. McGuire, Appl. Phys. 
Lett. \textbf{51}, 1040 (1987).

\bibitem{Bobroff99} J. Bobroff, W. A. MacFarlane, H. Alloul, P. Mendels, N. Blanchard, G. Collin,
 and J.-F. Marucco, Phys. Rev. Lett. \textbf{83}, 4381 (1999). 
\bibitem{Martinho} H. Martinho, A.A. Martin, C. Rettori, and C.T. Lin, Phys. Rev. B \textbf{69}, 180501(R) (2004).

\bibitem{com2} The fact that the pair breaking contributions correspond to the same energy in A$_{1g}$ and B$_{1g}$ 
channels can be explained by resonance effect as proposed in Ref. \onlinecite{Sherman02}.
\bibitem{Anderson} P. W. Anderson, Phys. Rev. \textit{112}, 1900 (1958).

\bibitem{Demler} E. Demler, Private communication.

\bibitem{Vesna} V.F. Mitrovic, E.E. Sigmund, M. Eschrig, H.N. Bachman, W.P. Halperin, A.P. Reyes, P. Kuhns, and W.G. Moulton,
Nature \textbf{413}, 501 (2001). 

\bibitem{Rossat91} J. Rossat-Mignod, L.P. Regnault, C. Vettier, P. Bourges, P. Burlet, J. Bossy, J. Y.
Henry and G. Lapertot, Physica C \textbf{185}, 86 (1991).

\end{references}
\end{document}